\documentclass[12pt]{article}

\newcommand{\sixthG}{\mathcal{G}_6}
\newcommand{\catalog}{\mathcal{C}}
\newcommand{\universe}{\mathcal{U}}

\renewcommand{\theequation}{\thesection.\arabic{equation}}

\title{\bf Spikes in cosmic crystallography II: topological signature 
of compact f\/lat universes}
\author{G.I. Gomero\thanks{E-mail: german@cbpf.br} , \  
M.J. Rebou\c{c}as\thanks{E-mail: reboucas@cbpf.br} , \
 \& A.F.F. Teixeira\thanks{E-mail: teixeira@cbpf.br} \\ 
\\ 
Centro Brasileiro de Pesquisas F\'\i sicas, \\
Rua Dr. Xavier Sigaud 150, \\
22290-180 Rio de Janeiro -- RJ, Brasil
}

\begin{document}
\maketitle

\begin{abstract}
We study the topological signature of euclidean isometries in 
pair separations histograms (PSH) and elucidate some unsettled 
issues regarding distance correlations between cosmic sources in 
cosmic crystallography. Reducing the noise of individual PSH's 
using mean pair separations histograms we show how to distinguish 
between topological and statistical spikes. We report results of 
simulations that evince that topological spikes are not enough to 
distinguish between manifolds with the same set of Clif\/ford 
translations in their covering groups, and that they are not the 
only signature of topology in PSH's corresponding to euclidean 
small universes. We also show how to evince the topological 
signature due to non-translational isometries.
\end{abstract}

\section{Introduction}
\label{intro}
\setcounter{equation}{0}

The method of Cosmic Crystallography (CC), devised by Lehoucq et 
al. \cite{LeLaLu}, looks for distance correlations between cosmic 
sources using pair separations histograms (PSH), i.e. plots of 
the number of pairs of sources versus the distance (or squared 
distance) between them. These correlations arise from the 
isometries of the covering group of the 3-manifold used to 
model our Universe and so they provide a signature of its global 
spatial topology. In this way CC is potentially useful to investigate 
the shape and size of our Universe. It has recently been shown by 
Gomero et al. \cite{Spikes} how to calculate the topological 
signature from these distance correlations in a very general 
geometrical-topological-observational setting. It turns out from 
the major result of Ref. \cite{Spikes} that correlations due to 
Clif\/ford translations manifest as spikes in PSH's, whereas other 
isometries manifest as small deformations of the \emph{expected} 
pair separations histogram (EPSH) of the corresponding universal 
covering manifold.

The major result of Ref.\cite{Spikes} has a striking consequence 
for universe models with hyperbolic spatial sections. Indeed, since 
no hyperbolic isometry is a Clif\/ford translation, there are no 
topological spikes in PSH's corresponding to low density universe 
models. Thus, at f\/irst sight, these histograms seem to give no 
reliable information of the topology of the spatial sections in 
these models of the universe. The absence of spikes in PSH's from 
hyperbolic universes is by now well understood and has been 
conf\/irmed by simulations performed by Lehoucq et al. \cite{LeLuUz} 
and Fagundes and Gausmann \cite{FagGaus2}. It remains, however, to 
understand the topological signature of hyperbolic isometries in CC.

The implications of the results of Ref.\cite{Spikes} for PSH's from 
f\/lat universe models seem to be less well understood. It has been 
stated in \cite{LeLuUz} and \cite{UzLeLu} that every euclidean 
isometry which produces $\Gamma$-pairs in a given catalog will give 
rise to a spike in the corresponding PSH. This statement, however, 
is in clear contradiction with the fact that only translations 
produce spikes~\cite{Spikes}. Moreover, in studying the applicability 
of CC to closed f\/lat models of our Universe, Fagundes and Gausmann 
\cite{FagGaus1} reported a PSH for a manifold of class 
$\sixthG$,\footnote{In this letter we use the notation of 
Ref.\cite{Wolf} to denote families of f\/lat compact orientable 
3-manifolds.} therein called model $E4$, which exhibits a 
signif\/icant peak at $(d/L)^2=5$. That paper suggests that this 
spike is generated by an isometry of the covering group of the 
manifold considered, and this interpretation was again suggested in 
Ref.\cite{FagGaus2}. Nevertheless, according to Ref.\cite{Spikes}, 
since there is no translation that would produce the peak at 
$(d/L)^2=5$, one immediately concludes that it must be due to 
statistical f\/luctuations, and so it is not of topological origin. 
A def\/initive elucidation of these unsettled issues would be 
useful because it would clarify the actual signature of euclidean 
non-translational isometries in PSH's. Indeed, by performing 
simulations we will evince in this letter that, contrary to what is 
suggested in \cite{LeLuUz} and \cite{UzLeLu}, topological 
spikes are not the only signature of topology in PSH's 
corresponding to euclidean small universes. Besides we also show 
through simulations that non-translational isometries do not 
manifest as \emph{less sharp peaks} as suggested by Fagundes and 
Gausmann \cite{FagGaus2}, but as broad and tiny deformations of 
the PSH corresponding to the simply connected case. Our results here 
are supported by and in agreement with the general theoretical 
developments of Ref.\cite{Spikes}.

Actually, the major purpose of this letter is to show how to use 
the MPSH technique described in \cite{Spikes} for studying the 
topological signature of isometries in PSH's. After a brief 
review of the techniques developed in \cite{Spikes} we 
f\/irst compute MPSH's for a manifold $M$ of class $\sixthG$ 
and reduce the statistical noise to a level that allows the 
identif\/ication of topological spikes. In this way (i) statistical 
spikes that may be confused with topological spikes are removed, 
and (ii) topological spikes that are masked by statistical 
f\/luctuations in individual PSH's show up even when there are 
few $\Gamma$-pairs corresponding to them. Incidentally, point 
(i) makes clear that actually there is no topological spike at 
$(d/L)^2=5$. As an additional application we construct an EPSH for 
the minimal 3-torus that covers $M$ and plot the dif\/ference 
between this EPSH and an MPSH of $M$. Since the covering groups of 
this 3-torus and that of $M$ have the same translations, then PSH's 
for these two manifolds would exhibit identical spike spectra 
\cite{Spikes}. So this dif\/ference yields the topological 
signature of non-translational isometries of the covering group of 
$M$ plus some statistical noise. For comparison, we also plot the 
dif\/ference between an MPSH and an EPSH, both for the 3-torus, 
obtaining as a result essentially statistical noise. This indicates 
that, within the accuracy of the simulations, topological spikes 
are the only topological signature in PSH's for a 3-torus.

\section{Pair separations histograms}
\label{histog}
\setcounter{equation}{0}

Here we brief\/ly review some results obtained in Ref.\cite{Spikes}, 
and extend them to the level needed for the development of this 
work. We begin by describing what a pair separations histogram 
(PSH) is, and then show how to construct mean pair separations 
histograms (MPSH) with simulated catalogs. We end with a brief 
explanation of the expected pair separations histogram (EPSH) 
and its use in determining the topological signature for 
non-translational isometries.

To build a PSH we simply evaluate a suitable one-to-one function 
$f$ of the separation $r$ of every pair of cosmic sources from a 
given catalog $\catalog$, and then count the number of pairs for 
which these values $f(r)$ lie within certain subintervals. These 
subintervals are all of equal length and must form a partition of 
the interval $(0,f(D)]$, where $D$ is the diameter of the observed 
universe corresponding to the catalog. A PSH is just a normalized 
plot of this counting. We will take the function $f$ to be the 
square function as is usual when dealing with f\/lat models in CC, 
in order to compare our plots with those found in the literature.

It is convenient to have a formal description of the above 
procedure. In considering discrete astrophysical sources, the 
\emph{observable universe} can be viewed as that part of the 
universal covering manifold $\widetilde{M}\,$ of the space-like 
section $M$ of spacetime, causally connected to an image of our 
position since the moment of matter-radiation decoupling; while, 
given a catalog of cosmic sources, the \emph{observed universe} 
${\universe}$ is that part of the observable universe which 
contains all the sources listed in the catalog. So, for instance, 
the observed universe corresponding to a catalog covering the 
entire sky is a ball of radius given by the redshift cutof\/f of 
the catalog, while for a pencil beam catalog it is a thin cone 
with vertex at an image of our position. Interesting observed 
universes that may be explored in the context of CC are thin 
spherical shells; these observed universes correspond to catalogs 
with approximately equal upper and lower redshift cutof\/fs 
\cite{Spikes}.

All the sources contained in ${\universe}$ can be observed in 
principle, but due to observational limitations a catalog 
consists only of part of them. Our observational limitations 
can be formulated as \emph{selection rules} which describe how 
the catalog arises from the set of observable images. These 
selection rules, together with the distribution law which the 
objects in $M$ obey, will be referred to as \emph{construction 
rules} for the catalog $\catalog$. It should be noted that the 
above def\/inition for a catalog f\/its in with the two basic types 
of catalogs one usually f\/inds in practice, namely real catalogs 
(which arise from observations) and simulated catalogs, which are 
generated under well-def\/ined assumptions that are posed to 
mimic some observational limitations and (or) to account for 
simplifying hypotheses.

To construct a PSH one begins by dividing the interval $(0,D^2]$ 
in $m$ equal subintervals of length $\Delta s = D^2/m$. Each 
subinterval has the form
\begin{displaymath}
J_i = \left( s_i - \frac{\Delta s}{2} \, , \, s_i + \frac{\Delta
s}{2} \right] \qquad ; \qquad i=1,2, \dots ,m \; ,
\end{displaymath}
and is centered at
\begin{displaymath}
s_i = \, \left( i - \frac{1}{2} \right) \,\, \Delta s \; .
\end{displaymath}
Given a catalog $\catalog$ of cosmic sources and denoting by 
$\eta(s)$ the number of pairs of sources in $\catalog$ with 
squared separation $s$, a PSH is then obtained plotting the 
function
\begin{equation}
\label{histograma}
\Phi(s_i)=\frac{2}{N(N-1)}\,\,\frac{1}{\Delta s}\,
               \sum_{s \in J_i} \eta(s) \; ,
\end{equation}
where $N$ is the number of sources in $\catalog$. Note that with 
the same catalog $\catalog$ we may obtain dif\/ferent PSH's 
simply by taking dif\/ferent values for $m$. The sum 
in~(\ref{histograma}) is just a counting and the coef\/f\/icient 
of the sum is a normalization constant, so 
\begin{equation}
\sum_{i=1}^m \Phi(s_i)\,\, \Delta s = 1 \, .
\end{equation}
Although it is usual in CC to refer to the plot of the function 
$\Phi(s_i)$ as a PSH, for theoretical purposes it is more useful 
to def\/ine a PSH as a function given by (\ref{histograma}). 
{}From now on we will always refer to $\Phi(s_i)$ simply as a PSH.

Any single PSH is plagued with statistical noise that may mask 
the topological signature. The simplest and most obvious way 
to reduce this noise is to use the MPSH which is described as 
follows. Consider $K$ comparable catalogs $\catalog_k$ 
($k=1,2,\dots,K$), with approximately the same number of 
cosmic sources and corresponding to the same manifold $M$. 
Let their PSH's, for a f\/ixed value of $m$, be given by
\begin{equation}
\label{sample-PSH}
\Phi_k(s_i) = \frac{2}{N_k(N_k-1)}\, \frac{1}{\Delta s}\,
\sum_{s \in J_i} \eta_k(s) \, ,
\end{equation}
where $N_k$ is the number of sources in $\catalog_k$ and 
$\eta_k(s)$ is the number of pairs of sources in $\catalog_k$ 
with squared separation $s$; then, the MPSH def\/ined by
\begin{equation}
<\!\Phi(s_i)\!>\,\, = \frac{1}{K} \,\sum_{k=1}^K \Phi_k(s_i)
\end{equation}
contains much less noise than any single PSH, and clearly 
contains the same topological information. Indeed, elementary 
statistics tells us that the statistical f\/luctuations in the 
MPSH are reduced by a factor proportional to $1/\sqrt{K}$, which 
makes at f\/irst sight the MPSH very attractive. In 
Sec.\ref{spikes} we apply this technique to discriminate between 
topological and statistical spikes in PSH's corresponding to an 
euclidean compact manifold.

As shown in Ref.~\cite{Spikes} (see also Refs.%
~\cite{GomRebTei2000,Signature}), in the 
limit $K \rightarrow \infty$ the MPSH approximates very well to the 
EPSH which is an ``ideal'' PSH, i.e. a PSH with the statistical 
noise completely removed. Equation~(4.15) of Ref.~\cite{Spikes} 
[or equivalently eq.~(2.11) rederived in Ref.~\cite{GomRebTei2000},
wherein $N= n(n-1)/2$ denotes the total number of pairs of cosmic
images] can be rewritten in the form
\begin{equation} \label{topsig1}
\Phi_{exp}\,(s_i) = \Phi^{sc}_{exp}\,(s_i) + \frac{\nu_u}{N-1}\, 
[\,\Phi^{u}_{exp}\,(s_i) - \Phi^{sc}_{exp}\,(s_i)\,] 
+ \frac{1}{N-1}\,\sum_{g \in \widetilde{\Gamma}} \nu_g\,
[\, \Phi^g_{exp}\,(s_i) - \Phi^{sc}_{exp}\,(s_i)\,] \;,
\end{equation}
where $\widetilde{\Gamma}$ is the covering 
group $\Gamma$ of $M$ without the identity element, $N$ is the 
mean value of the $N_k$, $\Phi_{exp}^{sc}(s_i)$ is the EPSH of 
the corresponding simply connected case, $N_u$ is the expected 
number of uncorrelated pairs and $\nu_u= 2\,N_u/N$. In~(\ref{topsig1}) 
$\Phi_{exp}^u(s_i) = F_u(s_i)/\Delta s$, where $F_u(s_i)$ 
is the probability of an uncorrelated pair to be separated by 
a squared distance that lies in $J_i$. 
For each covering isometry it is also def\/ined a number 
$\nu_g = N_g/N$, where $N_g$ is the expected number of $g$-pairs 
in a catalog with $N$ sources, and a distribution function 
$\Phi_{exp}^g(s_i) = F_g(s_i)/\Delta s$, where $F_g(s_i)$ 
is the probability of an observed $g$-pair to be separated by 
a squared distance that lies in $J_i$.

Now within the approximation\footnote{This approximation is 
justif\/ied \emph{a posteriori} by Fig.5a for the case of a 
3-torus.}
\begin{equation}
\label{approx}
\Phi_{exp}^{sc}(s_i) \approx \Phi_{exp}^u(s_i) \; ,
\end{equation}
the EPSH reads
\begin{equation}
\label{EPSH}
\Phi_{exp}(s_i) \approx \Phi_{exp}^{sc}(s_i) + \frac{1}{N-1}\, 
\sum_{g \in \widetilde{\Gamma}} \nu_g\, [\, \Phi_{exp}^g(s_i) -
\Phi_{exp}^{sc}(s_i)\,] \; .
\end{equation}

The general underlying setting for performing the calculations 
involved in (\ref{EPSH}) is the assumed existence of an ensemble 
of catalogs comparable to a given catalog $\catalog$ (real or 
simulated), with the same number of sources and corresponding to 
the same manifold $M$. The construction rules permit the 
computation of probabilities and expected values involved in 
(\ref{EPSH}).

Let $\Gamma_t \subset \Gamma$ be the subset of all Clif\/ford 
translations of $\Gamma$ (i.e. all the isometries $g \in \Gamma$ 
such that for all $p \in \widetilde{M}$, the distance $|g(p)|= 
d(p,gp)$ is independent of $p$). When $g \in \Gamma_t$ we have 
\begin{equation}
\label{spike}
\Phi^g_{exp}(s_i) = \frac{\delta_{i,i_g}}{\Delta s} \; ,
\end{equation}
where $\delta_{i,i_g}$ is the Kronecker delta, and $i_g$ is the 
position of the spike due to the translation $g \in \Gamma_t$, 
i.e. $|g|^2 \in J_{i_g}$. Then one can write (\ref{EPSH}) as
\begin{equation}
\label{EPSH2}
\Phi_{exp}(s_i) \approx \Phi_{exp}^{sc}(s_i) + \varphi_{exp}^t(s_i) + 
\varphi_{exp}^{nt}(s_i) \; ,
\end{equation}
where
\begin{equation}
\label{transEPSH}
\varphi_{exp}^t(s_i) = \frac{1}{N-1}\, \sum_{g\in \widetilde{\Gamma}_t} 
\nu_g\, [\, \frac{\delta_{i,i_g}}{\Delta s} - \Phi_{exp}^{sc}(s_i)\,] 
\end{equation}
is the contribution of Clif\/ford translations to the topological 
signature of the EPSH, and
\begin{equation}
\label{ntEPSH}
\varphi_{exp}^{nt}(s_i) = \frac{1}{N-1}\, \sum_{g\in \Gamma \setminus 
\Gamma_t} \nu_g\, [\, \Phi_{exp}^g(s_i) - \Phi_{exp}^{sc}(s_i)\,] 
\end{equation}
is the topological signature associated to the non-translational 
isometries of 
$\Gamma$.

It is clear from (\ref{EPSH2}) that manifolds with the same 
translations in their covering groups will exhibit the same 
spike spectra given by (\ref{transEPSH}), the only dif\/ference 
between their EPSH's being the topological signature associated to 
non-translational isometries. Moreover, in the euclidean case, from 
(\ref{EPSH2}) one can always write
\begin{equation}
\label{ntEPSH2}
\varphi_{exp}^{nt}(s_i) \approx \Phi_{exp}(s_i) - \Phi_{exp}^{torus}(s_i) 
\; ,
\end{equation}
with
\begin{equation}
\label{torusEPSH}
\Phi_{exp}^{torus}(s_i) \approx \Phi_{exp}^{sc}(s_i) + 
\varphi_{exp}^t(s_i) 
\end{equation}
being the EPSH of the minimal 3-torus that covers $M$.

Now, using the fact that $<\!\Phi(s_i)\!> \, \approx \Phi_{exp}(s_i)$ 
one gets another approximate expression for the topological signature 
of non-translational isometries of small f\/lat universes, namely
\begin{equation}
\label{ntEPSH3}
\varphi_{exp}^{nt}(s_i) \approx \, <\!\Phi(s_i)\!> - \; 
\Phi_{exp}^{torus}(s_i) \; .
\end{equation}
This expression can easily be numerically evaluated since the MPSH can 
be obtained with computer simulations, and $\Phi_{exp}^{torus}(s_i)$ 
is given explicitly by
\begin{equation}
\label{torusEPSH2}
\Phi_{exp}^{torus}(s_i) \approx \left( 1 - \frac{1}{N-1} \sum_{g \in 
\widetilde{\Gamma}_t} \nu_g \right) \Phi_{exp}^{sc}(s_i) + 
\frac{1}{\Delta s \, (N-1)} \sum_{g \in \widetilde{\Gamma}_t} 
\nu_g \delta_{i,i_g} \; .
\end{equation}
In Sec.\ref{TopSig} we use (\ref{ntEPSH3}) and (\ref{torusEPSH2}) to 
evince the shape of the topological signature of non-translational 
isometries for an euclidean closed manifold.

{}From now on we will consider only trivial construction rules, 
i.e. we will assume that cosmic sources are uniformly distributed 
in space, and all cosmic sources present in universe models, 
up to a given redshift, are recorded in catalogs. Although unrealistic, 
this assumption makes easy to illustrate the general results developed 
in Ref.\cite{Spikes} and permits a comparison with current literature 
in CC. Besides, in this case we can readily compute $\Phi^{sc}_{exp}(s_i)$ 
and the coef\/f\/icients $\nu_g$.

Indeed, from Ref.\cite{BerTei} (see also Ref.~\cite{Reboucas2000}) one can 
easily calculate 
$\Phi^{sc}_{exp}(s_i)$ for the case of trivial construction rules 
and a ball of radius $R$ as an observed universe. For f\/lat models 
one obtains
\begin{equation}
\label{scEPSH}
\Phi_{exp}^{sc}(s) = \frac{3}{32} \frac{\sqrt{s}}{R} \, (2 - 
\frac{\sqrt{s}}{R})^2 \, (4 + \frac{\sqrt{s}}{R}) \; \Theta(2 - 
\frac{\sqrt{s}}{R}) \; ,
\end{equation}
where $\Theta$ is the Heaviside function. Correspondingly, the 
coef\/f\/icients $\nu_g$ can be calculated by simple geometrical 
arguments. Indeed, let $\universe$ be the observed universe, i.e. 
a ball of radius $R$ centered at our position. The isometry $g$ 
transforms isometrically the ball $g^{-1}(\universe)$ into the ball 
$\universe$, so only the sources in $g^{-1}(\universe) \cap 
\universe$ have a $g$-partner in $\universe$, and form $g$-pairs. 
Thus we have
\begin{equation}
\nu_g = \frac{\mbox{Vol}(g^{-1}(\universe) \cap 
\universe)}{\mbox{Vol}(\universe)} 
\; .
\end{equation}
A simple calculation yields
\begin{equation}
\label{nug}
\nu_g = 1 - \frac{3}{4}\left(\frac{d_g}{R}\right) + 
\frac{1}{16}\left(\frac{d_g}{R}\right)^3 \; ,
\end{equation}
where $d_g$ is the distance from the center of the observed universe 
to its image by the isometry $g$.

\section{Topological and statistical spikes}
\label{spikes}
\setcounter{equation}{0}

Let us now show how to use the MPSH technique to discriminate 
between topological and statistical spikes by working out two 
examples of models of the universe reported in Ref.\cite{FagGaus1}. 
In order to make a comparison with the plots of the upper part 
of f\/ig.1 in Ref.\cite{FagGaus1}, we took a manifold $M$ of type 
$\sixthG$ with covering group $\Gamma$ generated by
\begin{eqnarray}
\label{generators}
\alpha(x,y,z) & = & (x+1,-y,-z) \, , \nonumber \\
\beta(x,y,z) & = & (-x,z+1,y) \, , \\
\delta(x,y,z) & = & (-x,z,y+1) \, , \nonumber
\end{eqnarray}
where $(x,y,z) \in R^3$. A fundamental polyhedron for $M$ and 
a detailed construction of this manifold is given in 
Ref.\cite{Gomero} (see also \cite{Oscillator}).

We have performed simulations for two observed universes with 
radii $R_1=\sqrt{2}$ and $R_2=5/6$, respectively. As in 
Ref.\cite{FagGaus1}, for each simulation in the f\/irst case 
($R_1=\sqrt{2}$) we put 20 objects uniformly distributed in the FP, 
while in the case $R_2=5/6$ we put 101 objects inside it. In both 
cases we end up with catalogs of approximately 240 sources. A PSH 
for one catalog for each case is shown in f\/ig.1, where 
we have subdivided the intervals $(0,4R_i^2]$ in bins of width 0.01 
as in Ref.\cite{FagGaus1}. A direct comparison with the plots in 
Ref.\cite{FagGaus1} can be done by considering unit lengths of 
$4200h^{-1}Mpc$ for the f\/irst case and $7200h^{-1}Mpc$ for the 
second. An important dif\/ference between our plots and those in 
Ref.\cite{FagGaus1} is that ours are normalized histograms as 
described in Sec.\ref{histog}. We can see that the PSH in 
f\/ig.1a presents no apparent spike at $s=5$, suggesting 
that the corresponding spike in Ref.\cite{FagGaus1} is of 
statistical origin. In order to elucidate this and related issues 
we shall now discuss the MPSH's corresponding to these PSH's.

{}Figure 2 shows one MPSH built with 50 computer generated 
comparable catalogs with approximately 240 sources for each case 
of study. It can be seen that, by taking the mean over 50 catalogs, 
the statistical noise has been considerably reduced. Moreover it 
became apparent from such plots that there is no topological spike 
at $s=5$ for the f\/irst model of the universe, whereas for the 
second there is a small spike at $s=2$ that is masked by statistical 
f\/luctuations in the individual PSH's of f\/ig.1b of this 
work and f\/ig.1 of Ref.\cite{FagGaus1}. Both MPSH's show the agreement 
between our simulations and the theoretical results of \cite{Spikes}. 
Besides, it can be shown that there are no topological spikes at any 
odd integer position in PSH's corresponding to models of the universe 
which have $M$ as spatial sections of spacetime (see Appendix A for 
details).

\section{Topological signature of non-translational isometries}
\label{TopSig}
\setcounter{equation}{0}

Let us now consider the group $\Gamma_t$ generated by the 
translations $a=\alpha^2$, $b=\beta^2$ and $c=\delta \beta^{-1}$. 
This group is formed by all the translations in $\Gamma$ and is 
the covering group of the minimal 3-torus that covers $M$. A 
fundamental polyhedron for this 3-torus is a parallelepiped of 
height 2 and square base of side $\sqrt{2}$. As it has been shown 
in Sec.\ref{histog}, PSH's corresponding to models of the universe 
that have the same translations in their covering groups exhibit 
the same spike spectra of topological origin. This result is 
illustrated in f\/ig.3 where it is shown a PSH, and 
an MPSH computed with 50 simulated catalogs in this 3-torus . All 
catalogs used here are comparable to those used in the simulations 
performed in Sec.\ref{spikes} for an observed universe of radius 
$R_1$, and have approximately 240 sources. We can note that there 
is no relevant dif\/ference between plots corresponding to $M$ 
and its minimal covering 3-torus even after reducing the noise 
by calculating the mean over 50 catalogs. Thus, looking at any 
of these plots one cannot say whether it comes from a catalog 
of sources in a $\sixthG$ universe or in its 3-torus minimal 
covering: topological spikes are not enough to distinguish 
between these two f\/lat manifolds because they have the same 
set of Clif\/ford translations in their covering groups.

To evince the topological signature of non-translational isometries 
of $M$ in CC, and so be able to distinguish between $M$ and any 
other manifold, we make use of (\ref{ntEPSH3}) and 
(\ref{torusEPSH2}). Fig.4 shows the graph of 
$\Phi_{exp}^{torus}(s_i)$ given by eq.(\ref{torusEPSH2}) 
together with~(\ref{scEPSH}) for the 
case of an observed universe of radius $R_1=\sqrt{2}$.
The comparison between the EPSH of f\/ig.4 and the MPSH shown
in f\/ig.3b makes apparent the strength of the MPSH procedure.
In both f\/ig.4 and f\/ig.5 we have reduced the number of 
sources from 240 to 120 to enhance the amplitude of the topological 
signature. In these f\/igures we have also incremented the width of 
the bins from 0.01 to 0.02 just to keep the amplitude of the 
spikes as in the previous simulations. Figs.5a and 5b are plots 
of eq.(\ref{ntEPSH3}). The f\/irst plot corresponds to 
eq.(\ref{ntEPSH3}) with an MPSH, $\,<\!\Phi(s_i)\!>\,$, for the 3-torus 
while the second corresponds to eq.(\ref{ntEPSH3}) with an MPSH for $M$. 
Both MPSH's were built with 5000 catalogs. Since the covering groups of 
this 3-torus and that of $M$ have the same translations, PSH's 
for these two manifolds would exhibit identical spike spectra; so 
the dif\/ferences given by eq.(\ref{ntEPSH3}) yield the topological 
signature of eventual non-translational isometries plus some 
statistical noise. Fig.5a is essentially noise, as expected, since 
there are no non-translational isometries in the covering group of 
the 3-torus; while f\/ig.5b evinces the topological signature of 
non-translational isometries of $\Gamma$ (the covering group of $M$). 
In f\/ig.5b, the amplitude of the topological signature of 
non-translational isometries is roughly one order of magnitude 
smaller than the amplitude and the statistical noise of the PSH's 
in f\/ig.1, so this signature could hardly be eventually extracted 
from a single PSH constructed with a real catalog. We have been 
able to evince this tiny topological signature in our simulations 
because we have succeeded in reducing the statistical noise by 
two orders of magnitude using the MPSH technique. Incidentally, 
if there is any other component to the topological signature in 
PSH's of 3-tori (i.e. other than spikes), its amplitude must be 
at least three orders of magnitude smaller than that of a PSH, 
as shown in f\/ig.5a.

One can obtain a better understanding of the structural features 
exhibited in f\/ig.5b, in particular the exact locations of the 
``jumps'', by noting that, since $\Gamma$ acts freely and 
discontinuously on euclidean space, there is a minimum non-null 
pair separation $|g|_{min}$ among all $g$-pairs, for each $g \in 
\widetilde{\Gamma}$. Thus, the terms $\Phi_{exp}^g(s_i)$, which 
appear in the def\/inition~(\ref{ntEPSH}) of the topological 
signature of non-translational isometries, are zero until 
the $i$-th bin given by the condition $|g|_{min}^2 \in J_i$. 
As reported in~\cite{Signature} (see fig.2a and fig.2b therein), 
these terms start with a jump at this $i$-th bin, therefore one 
expects that the topological signature of non-translational isometries 
presents such jumps at well def\/ined positions that we will 
calculate in what follows.

Note that all non-translational isometries of $\Gamma$ can be 
grouped in the following three categories according to 
whether the parameters $l$ and $m$ are even (E) or odd (O) (see 
Appendix A)
\begin{description}
\item[Type OE] $l=2k_1+1$ and $m=2k_2$ ,
\item[Type EO] $l=2k_1$ and $m=2k_2+1$ , and
\item[Type OO] $l=2k_1+1$ and $m=2k_2+1$ ,
\end{description}
where $k_1$ and $k_2$ are integers.

Straightforward calculations yield 
\begin{eqnarray*}
|\rho(2k_1+1,2k_2,n)|^2_{min} & = & (2k_1+1)^2 \, , \\
|\mu(2k_1+1,2k_2,n)|^2_{min} & = & (2k_1+1)^2 \, , 
\end{eqnarray*}
for type OE isometries;
\begin{eqnarray*}
|\rho(2k_1,2k_2+1,n)|^2_{min} & = & 2(k_2+n+1/2)^2 \, , \\
|\mu(2k_1,2k_2+1,n)|^2_{min} & = & 2(k_2+n+1/2)^2 \, , 
\end{eqnarray*}
for type EO isometries; and
\begin{eqnarray*}
|\rho(2k_1+1,2k_2+1,n)|^2_{min} & = & 2(n-1/2)^2 \, , \\
|\mu(2k_1+1,2k_2+1,n)|^2_{min} & = & 2(n+1/2)^2 \, .
\end{eqnarray*}
for type OO isometries. Here $\rho(l,m,n)$ and $\mu(l,m,n)$ are
given by equation~(\ref{canon}).

Now it can easily be seen that the non-translational contributions 
to the topological signature sharply start at $s_i=0.5,1,4.5,9,12.5 
\dots$ (see table~\ref{jumps} for clarif\/ication). An observable universe 
for a full sky covering survey and radius $R_1$ allows non-null 
contributions from non-translational isometries with 
$|g|^2_{min} < 4R_1^2 = 8$, and so its topological signature must present
``jumps'' at $s_i=0.5,1$ and $4.5$ only, in agreement with f\/ig.5b.

\begin{table}[t]
\begin{center}
\begin{tabular}{*{5}{|c}|} \hline
$|g|_{min}^2$ & \itshape Type & $k_1$ & $k_2+n$ & $n$ \\ \hline
\hline
     &      EO    &   --  & 0, -1 &   --  \\ \cline{2-5}
0.5  & OO$(\rho)$ &   --  &   --  &  0, 1 \\ \cline{2-5}
     & OO$(\mu)$  &   --  &   --  &  0, -1 \\ \hline
  1  &      OE    & 0, -1 &   --  &  --    \\ \hline
     &      EO    &  --   & 1, -2 &  --    \\ \cline{2-5}
4.5  & OO$(\rho)$ &  --   &   --  & -1, 2  \\ \cline{2-5}
     & OO$(\mu)$  &  --   &   --  &  1, -2 \\ \hline
  9  &      OE    & 1, -2 &   --  &   --   \\ \hline
     &      EO    &   --  & 2, -3 &   --   \\ \cline{2-5}
12.5 & OO$(\rho)$ &   --  &   --  & -2, 3  \\ \cline{2-5}
     & OO$(\mu)$  &   --  &   --  &  2, -3  \\ \hline
\end{tabular}
\end{center}
\caption{\label{jumps} The f\/irst f\/ive values for the squared 
minimum distance $|g|_{min}^2$ of $g$-pairs, and specif\/ications of 
the type of isometries together with the values of the corresponding 
parameters which give rise to them.}
\end{table}

To close this section we emphasize that the MPSH technique is a 
suitable approach to 
obtain the topological signatures only when one is dealing with 
simulated catalogs. Note however that the ultimate step in most 
of such statistical approaches to extract the topological 
signature is the comparison of the graphs (signature) obtained 
from simulated catalogs against similar graphs generated from 
real catalogs. To do so one clearly has to have the simulated 
patterns of the topological signatures of the manifolds, which 
can be achieved by the MPSH approach discussed in this work.

\section{Final remarks}
\label{dist-corr}
\setcounter{equation}{0}

In this letter we have studied, through simulations 
concerning closed f\/lat models of our Universe, some general 
results obtained in Ref.\cite{Spikes}, namely, that topological 
spikes in PSH's arise from translations, while 
non-translational isometries manifest as slight deformations of 
the EPSH of the corresponding simply connected case.

We have used the technique of taking means of PSH's, described in 
Ref.~\cite{Spikes}, to reduce the noise to a level that allows (i) 
to distinguish between topological and statistical spikes, and 
(ii) to evince the shape of the topological signature of 
non-translational isometries. In particular, we have shown 
explicitly that topological spikes for a cubic manifold $M$ of 
class $\sixthG$ with $L = 1$, appear only at even integers in plots 
of $n(d)$ vs. $d^2$, so the spike at $(d/L)^2 = 5$ reported in 
Fig.1 of Ref.\cite{FagGaus1} is indeed due to statistical 
f\/luctuations, in agreement with the theoretical conclusion we 
have derived from the results of \cite{Spikes}.

Identifying the shape of the topological signature of 
non-translational isometries is possible only through a drastic 
reduction of the statistical noise followed by the elimination of 
the uncorrelated part of the EPSH together with the topological 
spikes. This last step has been attained by subtracting  
the EPSH of the minimal 3-torus that covers $M$ from an MPSH built 
from $M$ and with
small enough noise. In doing so we have shown that the topological 
signature of non-translational isometries is formed by broad and 
tiny distributions, each one beginning with a ``jump''at 
$|g|^2_{min}$, for some isometry $g$.

The methods employed in this work allowed the study of the nature 
of the topological signature in CC of euclidean non-translational 
isometries, but can be equally applied to any of the other two 
geometries of constant curvature. Note, however, that these methods 
may not be useful in applications of CC to real catalogs, since 
it is impossible in practice to construct thousands, or even 
hundreds, of comparable catalogs of real cosmic sources. For an 
implementation of CC along these lines for use with real catalogs, 
a novel 
technique for reducing statistical noise with just one catalog 
has to be developed. We emphasize however that, despite the use of 
MPSH's be restricted to simulated catalogs, it is a suitable 
approach for studying the topological signature of non-translational 
isometries in PSH's, and thus it is a very important tool for the 
understanding of the method of CC. Clearly, without the 
understanding which arises from the MPSH approach to CC it would 
be quite dif\/f\/icult to establish the applicability of the 
crystallographic method as well as to devise 
alternative methods in a systematic way.

\section*{Acknowledgements}

We thank the Brazilian scientif\/ic agencies CNPq and CAPES for
f\/inancial support.

\section*{Appendix A}
\renewcommand{\theequation}{A.\arabic{equation}}

In this Appendix we show explicitly that there are no topological 
spikes at odd integer positions in PSH's constructed from models 
using $M$ (given by the covering group $\Gamma$ generated by 
(\ref{generators})) as spatial sections of spacetime. From the 
relations $\beta^2 = \delta^2$, $\beta \alpha = \alpha^{-1} \beta$, 
$\delta \alpha = \alpha^{-1} \delta$, and $\beta \alpha^{-1} = 
\alpha \delta^{-1}$, and from the fact that $\alpha^2$, $\beta^2$, 
$\beta \delta$ and $\delta \beta$ are translations, and so they 
commute, one can write any covering isometry of $M$ in one of the 
two canonical forms 

\parbox{11cm}{
\begin{eqnarray}
\rho(l,m,n) &=& \alpha^l \beta^m (\beta \delta)^n \, , \nonumber \\
\mu(l,m,n) &=& \alpha^l \beta^m (\delta \beta)^n \nonumber \, ,
\end{eqnarray} } \hfill
\parbox{1cm}{\begin{eqnarray} \label{canon} \end{eqnarray} }

\noindent
where $l$, $m$ and $n$ are integers, and $n \geq 0$. For all $l$ 
and $m$ we have that $\rho(l,m,0) = \mu(l,m,0)$, while for $n 
\neq 0$, $\rho(l,m,n) \neq \mu(l,m,n)$.

The translations in $\Gamma$ are those isometries with $l$ and 
$m$ even, regardless of the value of $n$. Explicitly we have
\begin{eqnarray*}
\rho(2k_1,2k_2,n)(x,y,z) = (x+2k_1,y+k_2+2n,z+k_2) \, , \\
\mu(2k_1,2k_2,n)(x,y,z) = (x+2k_1,y+k_2,z+k_2+2n) \, ,
\end{eqnarray*}
where $k_1$ and $k_2$ are integers, and so
\begin{equation}
\label{distance}
|\rho(2k_1,2k_2,n)|^2 = |\mu(2k_1,2k_2,n)|^2 = 4k_1^2+k_2^2+(k_2+2n)^2 
\, .
\end{equation}

Equation (\ref{distance}) gives the position of any potential 
topological spike in PSH's based on models of the universe with 
$M$ as a spatial section of spacetime. Whether a given spike will 
appear depends on the shape and size of the observed universe, 
i.e. on the shape and deepness of the region scanned by the 
astronomical survey used to construct the catalog. However,  by 
writing Eq.(\ref{distance}) in the form
\begin{equation}
\label{even-dist}
|\rho(2k_1,2k_2,n)|^2 = 2\left[2k_1^2+(k_2+n)^2+n^2\right] \, ,
\end{equation}
and an identical expression for $|\mu(2k_1,2k_2,n)|^2$, one can 
conclude that there are no spikes of topological origin at odd 
integer positions, since $|\rho(2k_1,2k_2,n)|^2$ and 
$|\mu(2k_1,2k_2,n)|^2$ take always even integer values. In particular, 
there is no topological spike at $s=5$. This is in agreement with the 
simulations which gave rise to f\/igs. 1 and 2, 
and shows that the spike found by Fagundes and Gausmann at 
$(d/L)^2=5$ is of statistical origin.

\vspace{1cm}
\section*{Captions for the figures}
\begin{description}
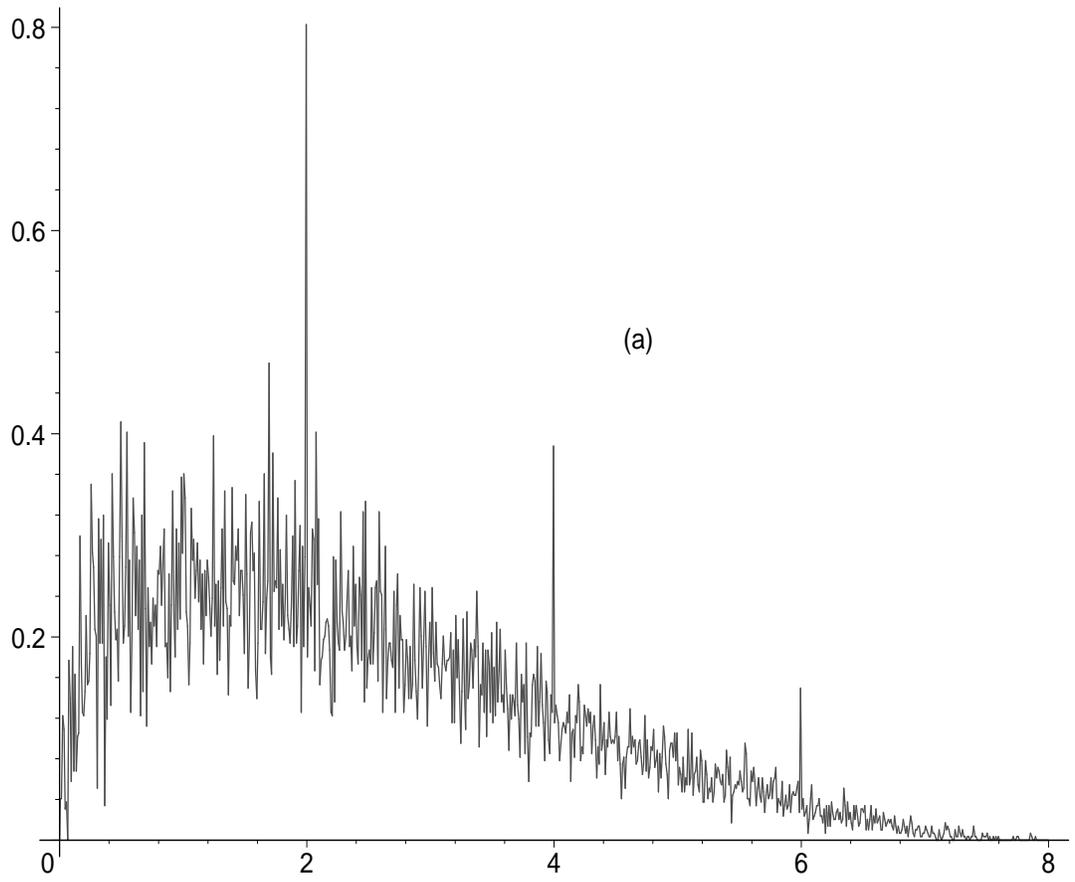
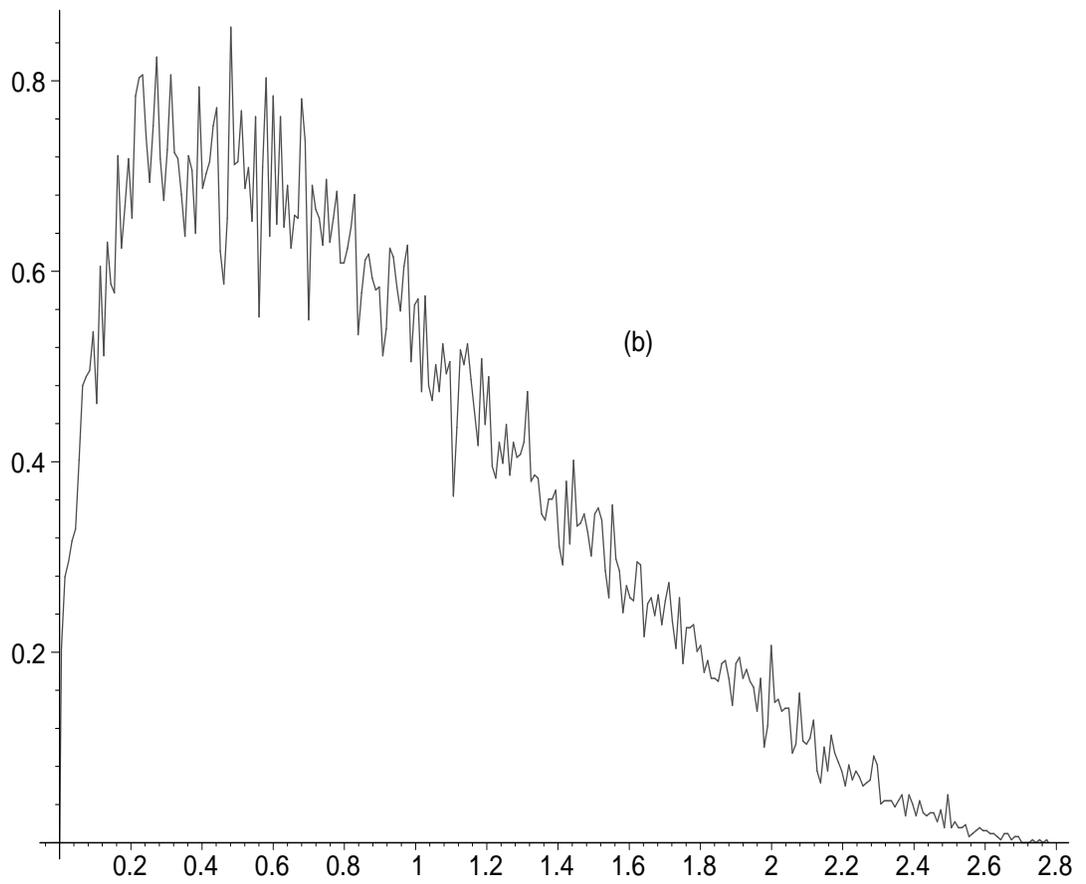
\item[Figure 1.] PSH's $\Phi(s_i)$ for two $\sixthG$ universes. 
In (a) the corresponding observed universe is a ball of radius 
$R_1=\sqrt{2}$, in (b) $R_2=5/6$. In both cases the catalogs have 
approximately 240 sources. The intervals $(0,4R^2]$ have been 
subdivided in bins of width 0.01. In (a) the PSH  presents no 
apparent spike at $s=5$.

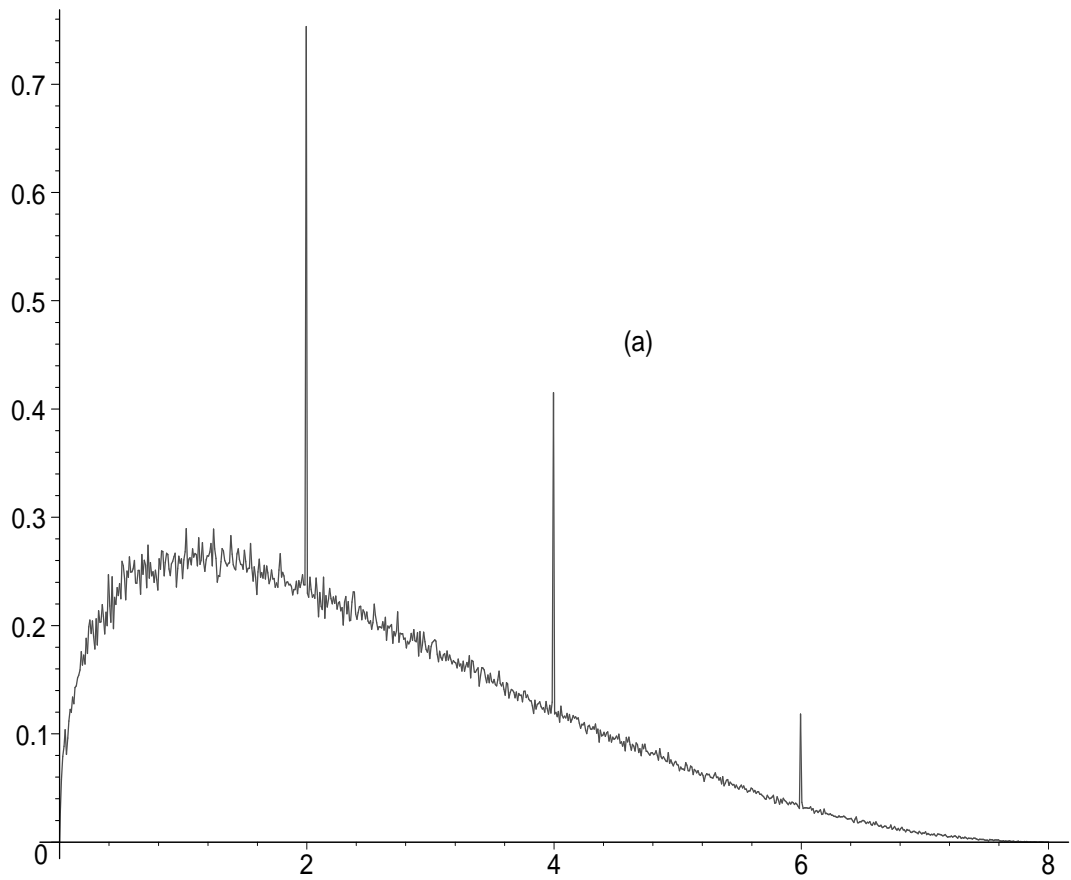
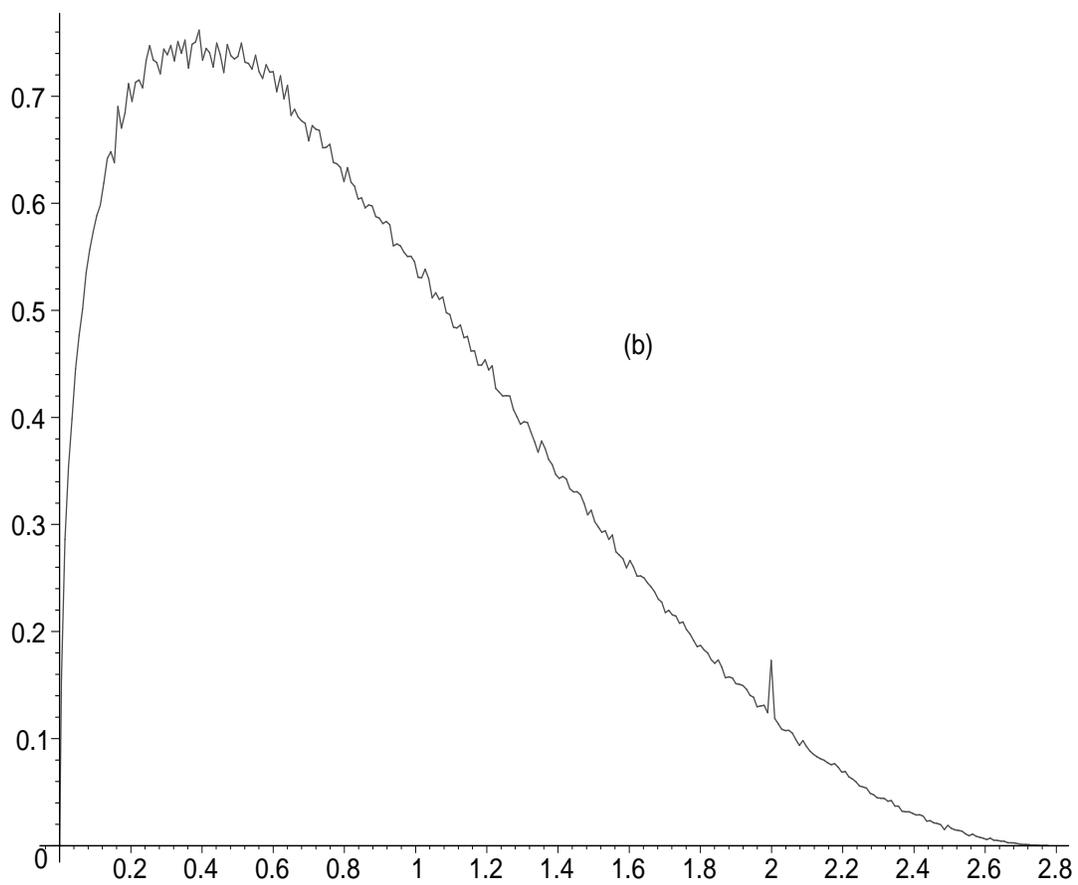
\item[Figure 2.] MPSH's $<\!\Phi(s_i)\!>$ built with 50 simulated 
comparable catalogs with approximately 240 sources for the two 
$\sixthG$ universes of f\/ig.1. The statistical 
noise has been considerably reduced so that it becomes apparent 
that there is no topological spike at $s=5$ in (a), whereas in 
(b) there is a small spike at $s=2$ that is masked by statistical 
f\/luctuations in the PSH of f\/ig.1b. The intervals 
$(0,4R^2]$ have been subdivided in bins of width 0.01.

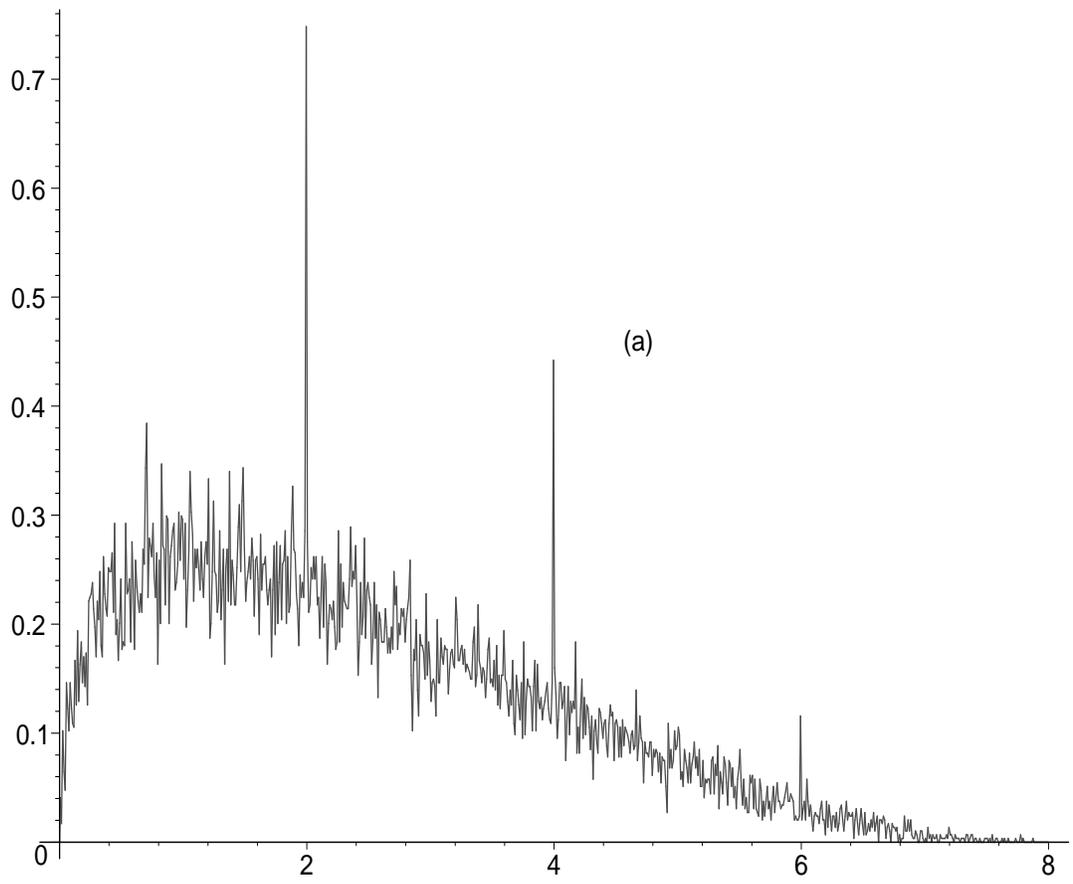
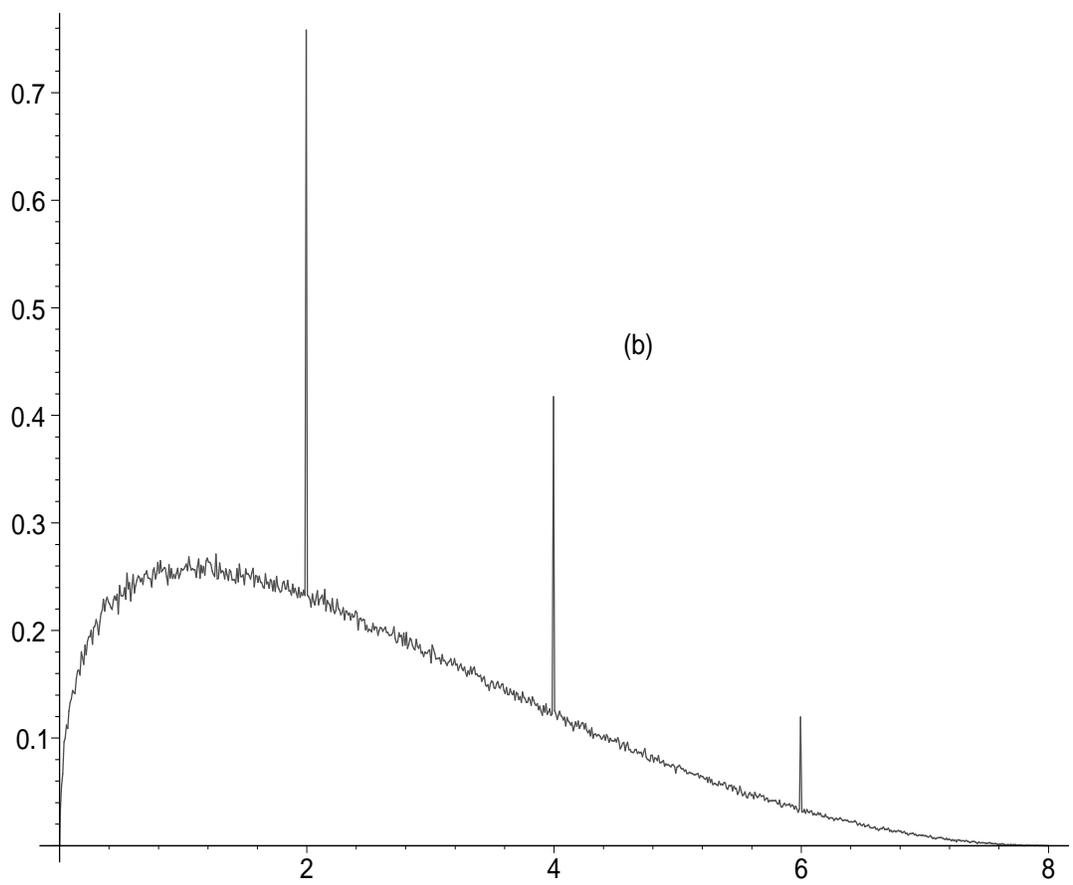
\item[Figure 3.] Part (a) is a PSH $\Phi(s_i)$ and part (b) is
a MPSH $<\!\Phi(s_i)\!>\,$ for the 3-torus universe which covers 
the $\sixthG$ universe of f\/igs.1a and 2a. All catalogs 
used here are comparable to those used for f\/igs.1a and 
2a, and have approximately 240 sources. The intervals 
$(0,4R^2]$ have been subdivided in bins of width 0.01. 
There is no relevant dif\/ference between graphs
corresponding to the $\sixthG$ model and its minimal covering 
3-torus, illustrating that topological spikes are not enough to 
distinguish between these two f\/lat manifolds.

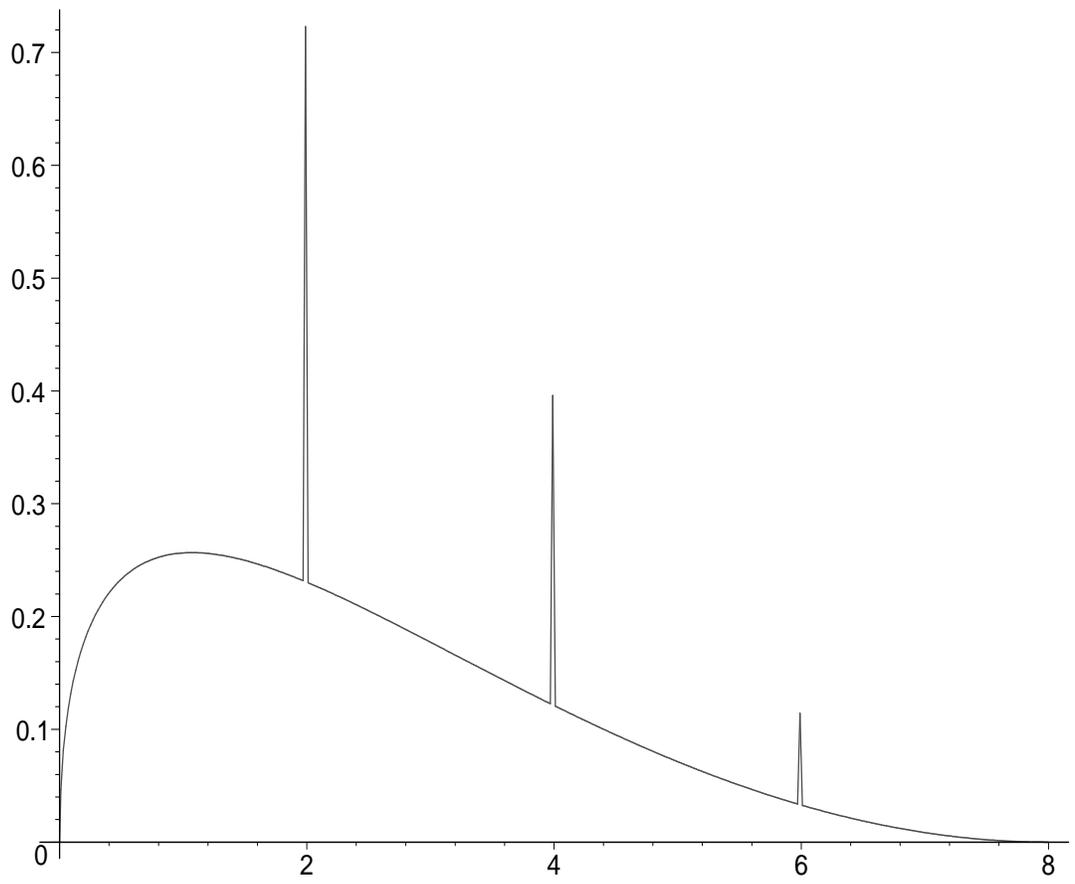
\item[Figure 4.] An EPSH $\Phi_{exp}(s_i)$ given by~(\ref{torusEPSH2}) 
for the 3-torus whose PSH $\Phi(s_i)$ is shown in f\/ig.3.    
The comparison between the EPSH of the present f\/igure with the 
MPSH of f\/ig.3b makes apparent the suitability and strength of the 
MPSH procedure.

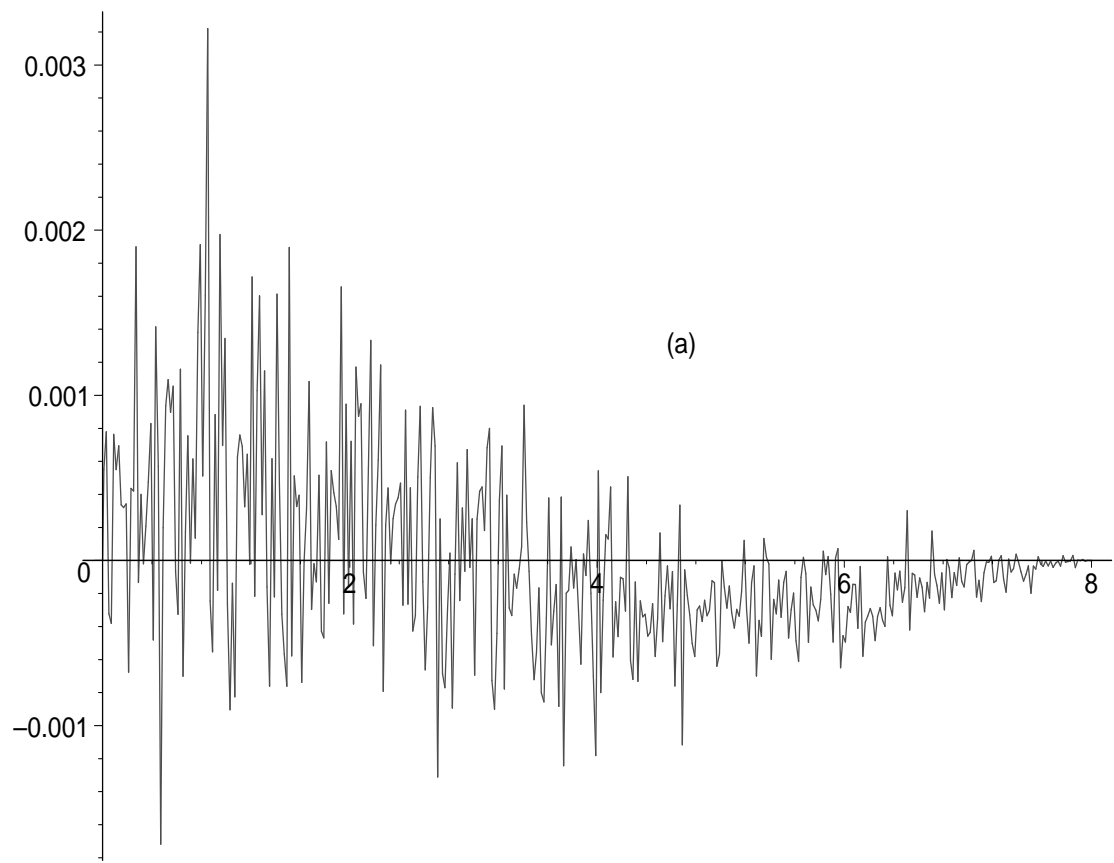
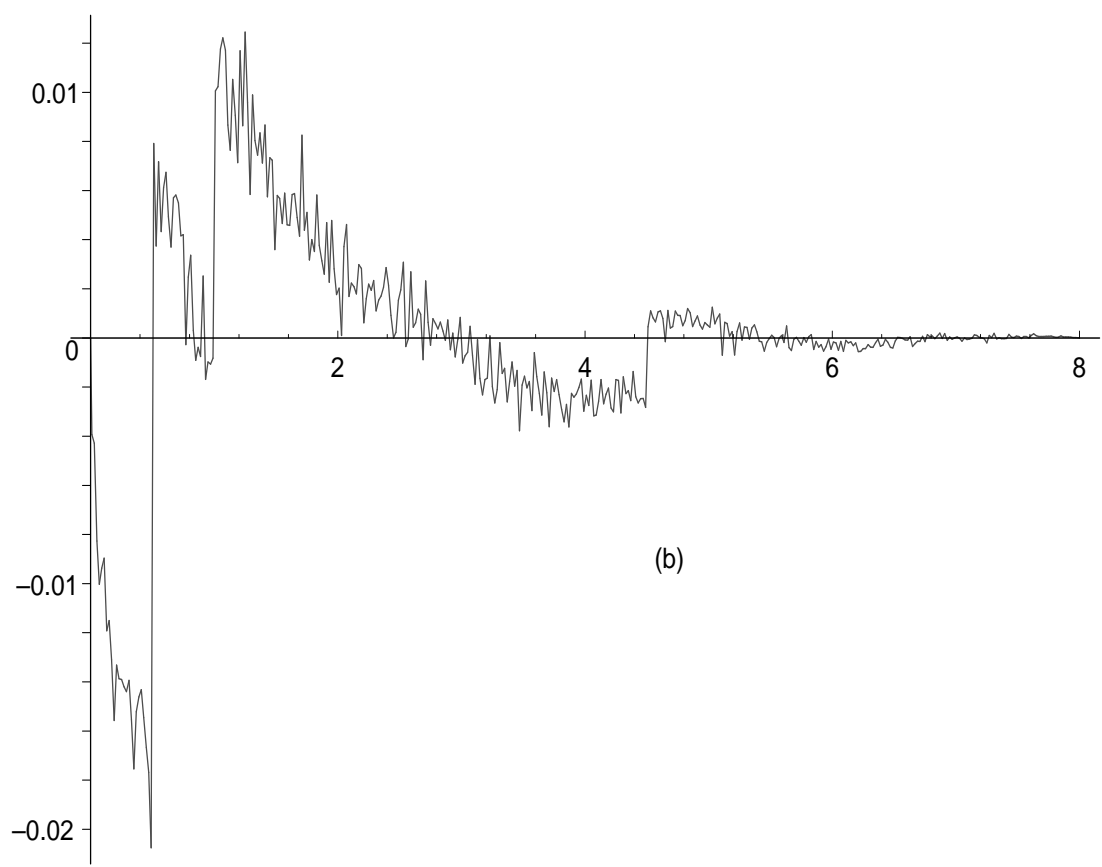
\item[Figure 5.] MPSH's corresponding to the topological signature of 
non-translational isometries given by (\ref{ntEPSH3}). Part (a) 
corresponds to the 3-torus of f\/ig.3 and f\/ig.4, while (b) 
corresponds to the $\sixthG$ manifold of f\/igs.1 and 2. 
Both MPSH's were built with 5000 catalogs of approximately 
120 sources, and with bins of width 0.02. While (a) exhibits 
essentially statistical noise as expected, (b) shows the 
topological signature of non-translational isometries 
of $\sixthG$. 

\end{description}

\end{document}